\documentstyle[twocolumn,aps,prl,epsf,floats]{revtex}
\newcommand{\la}[1]{\label{#1}}

\newcommand{\be}{\begin{equation}}
\newcommand{\ee}{\end{equation}}
\newcommand{\ba}{\begin{eqnarray}}
\newcommand{\ea}{\end{eqnarray}}
\newcommand{\bi}{\begin{itemize}}
\newcommand{\ei}{\end{itemize}}
\newcommand{\rmi}[1]{{\mbox{\scriptsize #1}}}

\newcommand{\nr}[1]{(\ref{#1})}

\newcommand{\tr}{{\rm Tr\,}}
\newcommand{\nn}{\nonumber}
\newcommand{\fr}[2]{{\frac{#1}{#2}}}
\newcommand{\msbar}{\overline{\mbox{\rm MS}}}
\newcommand{\lambdamsbar}{\Lambda_{\overline{\rm MS}}}
\newcommand{\dr}{{\rm dr}}

\def\lsi{\raise0.3ex
\hbox{$<$\kern-0.75em\raise-1.1ex\hbox{$\sim$}}}
\def\gsi{\raise0.3ex
\hbox{$>$\kern-0.75em\raise-1.1ex\hbox{$\sim$}}}
\newcommand{\lsim}{\mathop{\lsi}}
\newcommand{\gsim}{\mathop{\gsi}}

\begin{document}
\twocolumn[\hsize\textwidth\columnwidth\hsize\csname
@twocolumnfalse\endcsname

\draft
\title{Non-perturbative Debye mass in finite $T$ QCD}
\author{K. Kajantie$^{a,b}$,
M. Laine$^{c}$,
J. Peisa$^{d}$,
A. Rajantie$^{b}$, 
K. Rummukainen$^{e}$ and 
M. Shaposhnikov$^{a}$
}
\address{$^{a}$Theory Division, CERN, CH-1211 Geneva 23,
Switzerland}
\address{$^{b}$Department of Physics,
P.O.Box 9, 00014 University of Helsinki, Finland}
\address{$^{c}$Institut f\"ur Theoretische Physik,
Philosophenweg 16, D-69120 Heidelberg, Germany}
\address{$^{d}$Department of Mathematical Sciences,
University of Liverpool, Liverpool L69 3BX, UK}
\address{$^{e}$Fakult\"at f\"ur Physik, Postfach 100131, D-33501
Bielefeld, Germany}
\date{\today}
\maketitle

\vspace*{-5.0cm}
\noindent
\hfill \mbox{BI-TP 97/28, CERN-TH/97-191, hep-ph/9708207}
\vspace*{4.8cm}

\begin{abstract}\noindent
Employing a non-perturbative gauge invariant definition of the Debye
screening mass $m_D$ in the effective field theory approach to finite  
$T$ QCD,
we use 3d lattice simulations to determine the 
leading ${\cal O}(g^2)$ and to estimate the
next-to-leading ${\cal O}(g^3)$ corrections to $m_D$ in the high
temperature region. The ${\cal O}(g^2)$ correction is large 
and modifies qualitatively
the standard power-counting hierarchy picture of correlation lengths in
high temperature QCD.
\end{abstract}

\vspace*{0.2cm}

\vskip1.5pc]

QCD matter, a spatially and temporally extended system of matter
described by the laws of Quantum Chromodynamics, goes at high  
temperatures
into a quark-gluon plasma phase, in which color is no more confined
and chiral symmetry is restored. An essential quantity,
describing coherent static interactions in the plasma, is the inverse
screening length of color electric fields, the Debye mass $m_D$. The
Debye mass enters in many essential characteristics of static
properties of the plasma. Its numerical value is important for
phenomenological
discussions of formation of the quark-gluon plasma, for the analysis
of $J/\Psi$ and $\Upsilon$ suppression in heavy ion collisions, for the
computation of parton equilibration rates, etc. (see, e.g.
\cite{review}).

The definition and computation of the Debye mass for {\em abelian} QED
plasma is well understood \cite{fradkin}. The electromagnetic current
$j_\mu$ is a gauge-invariant quantity, and the Debye mass can be  
extracted
from the 2-point gauge invariant correlation function of $j_0$ in the
plasma. There are no massless charged particles in QED, which allows an
infrared-safe perturbative computation of the Debye mass in powers of
the electromagnetic coupling $e$. This has been done to order $e^5$
\cite{bip}. The situation in QCD is much more complicated. First, the
corresponding current in QCD, $j_\mu^a$, is not a gauge invariant
quantity. Second, there are massless charged gluons which give rise to
infrared divergences and prevent the perturbative determination of the
Debye mass beyond leading order.

A non-perturbative gauge invariant definition of the Debye mass in
vectorlike theories with zero chemical potential was suggested
in \cite{ay}. According to it, $m_D$ can be defined from the
large distance exponential fall-off of cor\-re\-la\-tors of  
gauge-invariant
time-reflection odd operators~$O$,
\be
\langle O(\tau,\vec{x})O(\tau,0) \rangle 
\sim C |\vec{x}|^\beta \exp(-m_D |\vec{x}|),
\ee
where $C$ and $\beta$ are some constants. The simplest choice for the
operator $O$ is $F^a_{03}F^a_{12}$, and other examples can be found in
\cite{ay}. In principle, 4-dimensional lattice simulations of hot QCD
would thus allow a measurement of the Debye mass at any temperature.

The aim of this letter is a non-perturbative determination of the {\em
high temperature limit} of the Debye mass, at $T > \mbox{a few}\times
T_c$.  We will see that the effective 3d approach to high
temperature gauge theories, developed in \cite{old,generic,effbn} (for
a review, see \cite{erice}) allows a simple and transparent gauge
invariant definition of the Debye mass~\cite{ay}, while 3d lattice
Monte Carlo simulations provide an economical way to determine its
value. The corrections to the leading result we shall find are
numerically large; thus many computations in the phenomenology of
quark-gluon plasma in heavy ion collisions should be re-analysed.

The theory we shall study is QCD with $N_f$ massless quark
flavours and with the gauge group SU($N$) with $N=2,3$. 
At high temperatures and
zero chemical potential the Debye mass can be expanded in a power  
series
in the QCD coupling constant $g=g(\mu)$ (the scale $\mu$ will be
specified later; the result for $N_f=0$ is shown explicitly in
eq.~\nr{m3}):
\ba
m_D & = & m_D^\rmi{LO}+{Ng^2T\over4\pi}\ln{m_D^\rmi{LO}\over  
g^2T}\nonumber\\
 & + & 
c_N g^2T + d_{N,N_f} g^3 T + {\cal O}(g^4T).
\label{md4d}
\ea
The leading order (LO) perturbative result, $m_D^\rmi{LO}
=(N/3+N_f/6)^{1/2}gT$, has been known for a long time~\cite{mDearly}.  
The
logarithmic part of the ${\cal O}(g^2)$ correction 
can be extracted perturbatively
\cite{rebhan}, but $c_N$ and the higher order corrections are 
non-perturbative. We are going to evaluate 
numerically the coefficients $c_N$ and
$d_{N,N_f}$. 

Static Green's functions for bosonic fields of high temperature QCD at
distances $|x| \gg T^{-1}$ we are interested in can be determined by
constructing an effective 3d gauge theory, containing static
magnetic gluons and the zero component of the 4d gauge field, $A_0$
\cite{old,generic,effbn}. Moreover, a 
{\em super-renormalizable} 3d theory,
defined by the Lagrangian 
\ba
L_\rmi{eff}[A_i^a,A_0^a] & = &  \fr14  F_{ij}^aF_{ij}^a
+ \tr [D_i,A_0][D_i,A_0]  \nn \\
& + & m_3^2\tr A_0^2 +\lambda_A(\tr A_0^2)^2,
\la{leff}
\ea
gives the Green's functions to a relative accuracy 
${\cal O}(g^4)$~\cite{generic}, 
which is sufficient for the accuracy of the expansion
in eq.~\nr{md4d}. The parameters of the effective theory are related to the
parameters of 4d QCD ($\lambdamsbar, N,N_f$) and 
the temperature as described in \cite{adjoint2}. For brevity,
we give here the explicit expressions only for $N_f=0$:
\ba
g_3^2 &=& g^2(4\pi e^{-\gamma_E
-\frac{1}{22}} T)T={24\pi^2 T\over 11N
\ln(6.742T/\lambdamsbar)},\la{g3}  \\
m_3^2&=&\fr{N}3 g^2(4\pi e^{-\gamma_E -\frac{5}{22}} 
T)T^2, \la{m3}\\
\lambda_A&=&{6+N\over24\pi^2}g^4(4\pi e^{-\gamma_E -\frac{7}{44}}
T)T. \la{lambdaA}
\ea
Here $g(\mu)$ is the QCD coupling in the $\msbar$ scheme and all the
effective theory couplings have been computed including both the 
leading and the next-to-leading order contributions.
The couplings \nr{g3}-\nr{lambdaA} are independent of the gauge 
chosen for the perturbative computation. The expansion parameter
is $g^2/16\pi^2$ so that the result should be accurate down to
$T\approx \mbox{\rm a few}\times \lambdamsbar$.

The dynamics of the 3d effective theory is fully characterised by 
the two dimensionless ratios
\be
y = {m_3^2\over g_3^4},\qquad x={\lambda_A\over g_3^2},
\la{def}
\ee
and by the dimensionful coupling $g_3^2$.  
The value of $x$ is essentially fixed by $T$,
\ba
x & = & {6+N\over24\pi^2}g^2(4\pi e^{-\gamma_E-3/11}T) 
\nonumber \\
& = & 
{6+N\over11N}\,\,{1\over\ln(5.371T/\lambdamsbar)},
\la{x}
\ea
while  $y$ and $x$, corresponding to physical 4d finite $T$ QCD for
$N_f=0$, are related by:
\ba
y = y_\dr(x)&=& {2\over9\pi^2x}+{1\over4\pi^2}+{\cal O}(x),
\quad N=2\la{ydrntwo}\\
&=&{3\over8\pi^2x}+{9\over16\pi^2}+{\cal{O}}(x),\quad N=3.
\la{ydrnthree}
\ea

We are now ready to give a gauge-invariant definition of the Debye mass
in the 3d language~\cite{ay}. 
Physically, we want a local operator which makes
$A_0^a$ gauge invariant in the 3d theory and contains $A_0^a$ singly.
We can single out this state by a symmetry consideration. Note that the
effective Lagrangian (\ref{leff}) has a discrete symmetry $A_0
\leftrightarrow -A_0$. Then the Debye mass 
can be defined as the mass of the lightest 3d state
which is odd under this symmetry. The
operator of lowest dimension of this type is 
\be
 h_i=\epsilon_{ijk}\tr A_0F_{jk}\,. 
\ee
Here the field $A_0^a$ has been made gauge invariant by
dressing it with a cloud of magnetic gluons.

At high $T$, one has $g\ll1$ and,
according to
eqs.~\nr{g3}, \nr{m3}, $m_3\gg g_3^2$. This is the ``heavy
quark limit'' of the 3d theory, in which the mass $m_D$ of the singlet
state is dominated by the bare mass $m_3$ of the scalar ``quark''
$A_0^a$. For dimensional reasons, the exact mass $m_D$ can in this
limit be expanded as
\be
 m_D= m_3 + a_N g_3^2 + 
\frac{b_N g_3^4}{m_3} +{\cal O}(\lambda_A,{g_3^2\lambda_A\over m_3},
{g_3^6\over m_3^2},\ldots),
\label{md3d}
\ee
where $a_N$ and $b_N$ are constants, perhaps involving a logarithm of
$m_3/g_3^2$. The terms neglected are of higher order using the
power counting in eqs.~\nr{g3}-\nr{lambdaA}.
Comparing eqs.~(\ref{md4d}) and (\ref{md3d}), one sees that 
\ba
a_N & = & {N\over4\pi}\ln{\sqrt{N/3+N_f/6}\over g}+
c_N,\nonumber \\
d_{N,N_f} & = &  \frac{b_N }{\sqrt{N/3+N_f/6}}.
\ea
Here we used the fact that the scale dependence of the 
non-perturbative terms in eq.~\nr{md4d} is at least of order ${\cal  
O}(g^4)$.
Since the expansion \nr{md3d} refers only to the 3d theory, the
constants $a_N$ and $b_N$ depend on $N$  but clearly not on $N_f$. Thus
$c_N$ is $N_f$ independent, while $d_{N,N_f}$ depends on $N_f$ only  
through
$m_D^\rmi{LO}$. 

In terms of our dimensionless variables 
\nr{def}, eq.~\nr{md3d} becomes
\be
\frac{m_D}{g_3^2}=\sqrt{y}+
\frac{N}{4 \pi} \ln\sqrt{y} + c_N  +  \frac{b_N}{\sqrt{y}} +
\ldots
\label{cconst}
\ee
\begin{figure}[ht]




\epsfysize=14cm
\centerline{\epsffile{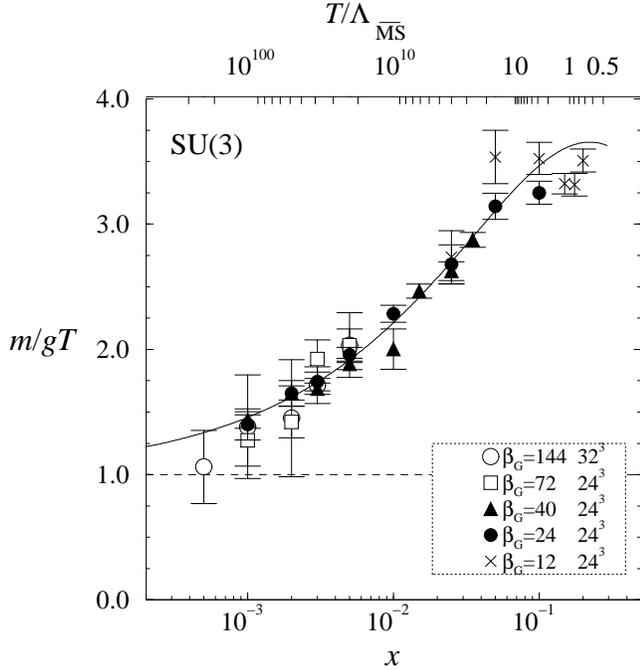}}

\vspace*{-5cm}

\caption[a]{The gauge invariant Debye mass 
for SU(3), as a function of $x$, or
$T/\lambdamsbar$ through eq.~\nr{x}.
The scale of $g$ on the $y$-axis is fixed according to eq.~\nr{m3}.
The dashed line marks the leading order $m_D^\rmi{LO}$, and the continuous
line the 2-parameter fit to eq.~\nr{md3d} with the parameters as in 
eq.~\nr{parameters}.}
\la{masses}
\end{figure}

The mass $m_D$ can now be measured by putting the effective 3d theory 
on the lattice \cite{adjoint2} and by
measuring the exponential falloff of the correlator
$\langle h_3(x_3)h_3(0)\rangle\sim\exp(-m_D|x_3|)$, where 
$h_3(x_3)$ is summed over the transverse ($x_1,x_2$) plane.
The correlation function is measured both with zero and finite
transverse momentum, and in order to enhance the overlap with the 
asymptotic state the measurements are performed with several
levels of recursive blocking of the operators.
We select the blocking level and momentum sector which
has the best signal for the asymptotic mass separately 
for each Monte Carlo run.
Since the longest correlation length in each case is less than
$1/5$ of the linear size of the lattice, we expect the finite volume
effects to be negligible
in comparison with the statistical errors.  This was also explicitly
checked by performing simulations with different volumes in isolated
cases.

The mass $m_D$ is defined in the whole $y,x$ parameter space.
To have results which are relevant for 4d physics, we perform the
measurements along the 2-loop dimensional reduction lines $y_\dr(x)$,
eqs.~(\ref{ydrntwo}--\ref{ydrnthree}).  To
measure the coefficients in eq.~\nr{cconst} one should use the part
of this curve corresponding to $\sqrt{y}\gg1$.  The results 
for $N=3$ are shown in
Fig.~\ref{masses}, in units of 4d $gT$ ($ = g_3^2\sqrt{3y/N}$ in 3d
units).   The Monte Carlo runs are 
performed with several lattice spacings $a$, parametrised by $\beta_G
\equiv 2N/(g_3^2 a)$.  For SU(3) $\beta_G$ varies by more than an
order of magnitude (although not at the same value of $x$), as shown
in Fig.~\ref{masses}; for SU(2), the measurements are done with
$\beta_G = 20$ and 32.   
The top scale of Fig.~\ref{masses} shows the physical temperature
$T/\lambdamsbar$ along $y_\dr(x)$ -line.  Note that the highest
temperatures are larger than $10^{100}\times \lambdamsbar \sim
10^{100} \times T_c$.

At small $x$ (large $y$), the fit to the function \nr{md3d} is very
good, as indicated by the continuous line in Fig.~\ref{masses}.  In
order to see in detail the sensitivity of the fit to the parameters,
in Fig.~\ref{fit} we replot the SU(3) data (restricted to $x<0.05$) in
terms of the quantity $\delta m/g_3^2 = {m_D}/{g_3^2}-\sqrt{y}-
\frac{N}{4 \pi} \ln\sqrt{y}$ as a function of $1/\sqrt{y}$.  The
intersection of the curve with the vertical axis gives the value of
$c_N$ and the slope gives $b_N = d_{N,N_f}\sqrt{N/3+N_f/6}$.  One can
see that the linear fit is rather good even down to small values of
$\sqrt{y}$.  The large non-zero value
of the intercept is very robustly determined. The slope $b_N$ is 
small and has a relatively large error. Only the statistical error is
given, but the value of $b_N$ also depends on the range of $y^{-1/2}$ 
included.

\begin{figure}[t]

\vspace*{-1.0cm}

\hspace{1cm}
\epsfysize=14cm
\centerline{\epsffile{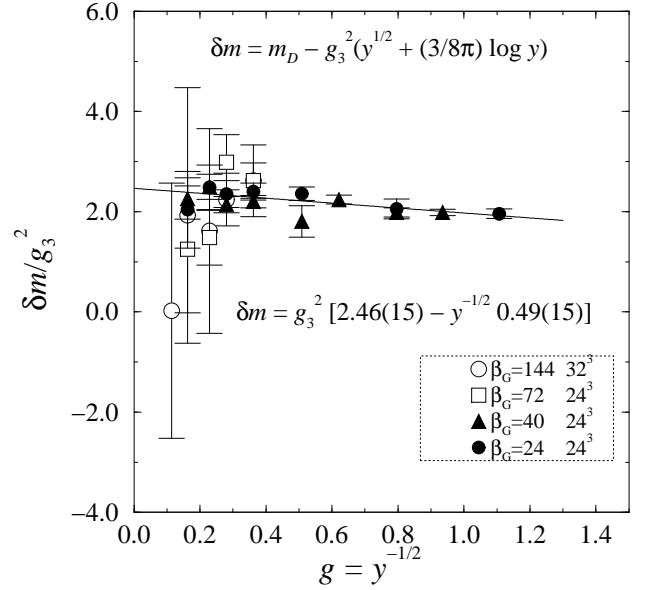}}

\vspace*{-4.6cm}

\caption[a]{The corrections of order ${\cal O}(g_3^2)$ and ${\cal
O}(g_3^2/\sqrt{y})$ to the SU(3) screening mass $m_D$ in 3d units,
corresponding to eq.~\nr{cconst}.}  \la{fit}
\end{figure}

The result of the fits are
\be\begin{array}{cll}
\mbox{SU(2):} & c_N = 1.58\pm 0.20\,\, & b_N =-0.03\pm 0.25 \\
\mbox{SU(3):} & c_N = 2.46\pm 0.15\,\, & b_N =-0.49\pm 0.15 \,.
\end{array}
\la{parameters}
\ee
The large number $c_N$ is related to non-perturbative 3d effects,
while the smaller $d_{N,N_f}$ can be viewed as being related to the
choice of scale in $m_D^{\rm LO}$.  
For $N=2$ we can in practice only verify that $d_{2,N_f}$ is close to zero.
Note that writing $c_N=N\tilde
c_N$, one has $\tilde c_N =0.79\pm 0.10$ ($N=2$), $0.82\pm 0.05$
($N=3$).

One can observe the following:
\bi
\item The leading term is dominant only at extremely large $T$. For
SU(3), the leading term is larger than the ${\cal O}(g^2)$
correction for $g<1/2.46$ or for
$T/\lambdamsbar\gsim
\exp(8\pi^22.46^2/11)\gsim 10^{19}$. This implies that the leading term
only dominates when QCD anyway merges into a unified theory.
\item The four terms in eq.~\nr{cconst} fit the data over all the range
$T\gsim 100T_c$
rather well, and there is no need for further corrections.
\item In the range $\lambdamsbar\approx T_c\lsim T\lsim 100T_c$,
$m_D$ is rather
constant and $\approx3.0\,m_D^{\rm LO}$ for SU(2) and
$\approx3.3\,m_D^{\rm LO}$ for SU(3). It should be noted, though, 
that in this regime $m_D^{\rm LO} > T$ so that the hierarchy
$m_D^2/(2\pi T)^2\ll 1$ required for an accurate description of 
4d physics through a 3d effective theory is getting weaker.
\item The mass measured from the $\langle A_0A_0\rangle$-correlator 
in the Landau gauge in 4d simulations
for $N_f=0$ has also been observed to be clearly larger
than the leading term \cite{karsch}.
\item  If the mass $\sim gT$ of the $A_0^a$ field is ``large'',
larger than the non-perturbative ${\cal O}(g^2)T$ correction,
the $A_0^a$ field can be further perturbatively integrated out
and a simpler effective theory, containing only $A_i^a$ (and
possible scalar fields) can be derived \cite{old,generic,effbn}.
Our results imply that this can be accurately carried out for
QCD only at extremely high temperatures, $T\gg T_c$. 
In the electroweak case the accuracy of the integration is sufficient
even for $T\sim T_c$ both since the leading term has a bigger
coefficient ($m_D^\rmi{LO}=\sqrt{11/6}gT$) than the $N_f=0$ QCD
considered here and 
since in the relevant $T$ regime the coupling constant 
$g=g(m_W)\approx2/3$ is smaller. 
\item The usual parametric ``power counting'' picture of correlation
lengths in high temperature QCD says that the longest scale, related
to the magnetic sector of the theory, is $m_M^{-1}\sim({\rm
const}\times g^2T)^{-1}$. A shorter scale, $\sim (g T)^{-1}$, is
associated with Debye screening. Our results show that this picture
can be quantitatively correct only at extremely large
temperatures. Indeed, purely magnetic effects, as measured by the
3d glueball (operator $F_{ij}^aF_{ij}^a$) mass ($m_G \approx 2 g_3^2$
for pure SU(2) \cite{phtw,adjoint2}) 
tend to be numerically
large, so that $m_M \sim m_D$ in a very wide range of temperatures 
(This gauge invariant result is in contrast to the small magnetic
gluon masses measured in Landau gauge \cite{karsch}).
In this range the
longest length scale corresponds to a scalar $0^{++}$ 3d ``bound
state'' of two $A_0$ quanta, associated with the operator $A_0^aA_0^a$
(the power counting suggestion that this state is roughly twice
as heavy as $m_D$ holds again only at extremely high $T$).
\end{itemize}

Summarizing, we have carried out with lattice Monte Carlo techniques a
gauge independent measurement of the Debye mass in finite $T$ QCD. The
measurement is based on first deriving with 2-loop perturbative
computations a 3d effective theory. The expansion parameter is
$\alpha_s/\pi$, so that the result is accurate down to $T$ close to
$T_c$. The mass is obtained by measuring correlators of the gauge
invariant local operator $A_0^aF_{jk}^a$ in the 3d theory. The leading
and next-to-leading corrections to $m_D$ were determined and found to
be large. In fact, for temperatures from $T_c$ up to $T \sim
1000\lambdamsbar$ the non-perturbative Debye screening mass is about
a factor $3$ larger than the lowest order estimate.

It remains to be seen whether this modification of the standard picture
of high-temperature gauge theories has applications in the
cosmological discussion of the quark-hadron phase transition or in
the phenomenology of heavy ion collisions.

The simulations were carried out with a Cray C94 and Cray 
T3E at the Finnish Center for Scientific Computing.

\end{document}